\documentstyle[seceq,epsf,preprint]{ptptex}

\newcommand{\bra}[1]{\langle {#1} |}     
\newcommand{\ket}[1]{| {#1} \rangle}     
\newcommand{\wtilde}[1]{\widetilde{#1}} 

\title{
Fermionic Squeezed State for Simple Algebraic Models in Many-Fermion Systems }

\author{
Hideaki {\sc Akaike},$^{1}$ 
Yasuhiko {\sc Tsue}$^{2}$ and Seiya {\sc Nishiyama}$^{2}$
}

\inst{
$^1$Department of Applied Science, Kochi University, Kochi 780-8520, Japan\\
$^{2}$Physics Division, Faculty of Science, Kochi University, Kochi 780-8520, 
Japan
}

\recdate{
\today
}

\abst{
It is shown that a variational approach
with fermionic squeezed states to many-fermion systems such as
pairing model is one of useful methods beyond the usual
Hartree-Fock-Bogoliubov approximation.
A pairing-type quasi-spin squeezed state is constructed
and adopted as a trial state in the variational method.
By using this state, quantum fluctuations are taken into
account.
In pairing model, dynamical and static approaches
are discussed in the context of comparing with the Hartree-Fock-Bogoliubov 
approximation about ground state energy. As a possible extension
to the $O(4)$ model, prospective squeezed states are
investigated and the usefulness is discussed.
}

\begin{document}

\maketitle

\section{Introduction}
As a possible extension of the time-dependent Hartree-Fock (TDHF) theory, 
a quasi-spin squeezed state was introduced instead of the Slater determinantal 
state as a possible extension of the trial state of 
variation.\cite{TKY94,TAKY96} 
This state realizes the minimum uncertainty relation as well as the Slater 
determinantal state as a certain kind of coherent state. 
However, contrary with 
the coherent state approach, the quasi-spin squeezed state 
approach can take account of degrees of freedom for quantum fluctuations 
dynamically. 
Hereby, the squeezed state approach gives the better approximation than 
the coherent state approach in general. 

In the preceding work,\cite{TKY94,TAKY96} applications of 
the quasi-spin squeezed 
state for many-fermion systems carried out in simple algebraic models. 
For example, the application to the Lipkin model\cite{LMG65} 
where the particle-hole 
interaction is active shows clearly that the quasi-spin squeezed state 
gives the useful and powerful approximation,\cite{TAKY96} 
because quantum fluctuations 
are contained properly. 
In this model, the quasi-spin squeezed state was constructed so as to 
take account of the particle-hole correlation. We call here 
it the particle-hole 
type squeezed state. This state reproduced the phase transition neatly, 
and in a certain limiting case for this state, the RPA equation was 
recaptured.\cite{TAKY98,KPTY01} 
From this viewpoint, the variational method with the quasi-spin squeezed state 
surpasses the RPA for many-fermion systems.

This paper is devoted to the purpose of the application and promotion 
of the quasi-spin squeezed state to the models with pairing interaction. 
One of the present authors (Y.T.) with Yamamura and Kuriyama have given 
a way to construct the quasi-spin squeezed state for pairing model.\cite{TKY94}
Also, two of the present authors (H.A. and Y.T.) have given a possible 
treatment of the degree of freedom for the quantum fluctuations in the 
Lipkin model.\cite{TA03}
In this paper, we extend the treatment previously given\cite{TA03} to 
the dynamical treatment in the pairing model. 
In the pairing model, we investigate the ground state energy with two 
different 
manners 
using the quasi-spin squeezed state. One is that the quasi-spin squeezed 
state is used in the framework of the time-dependent variational principle, 
and the system is considered dynamically. 
The other is that the energy minimum is sought directly introducing the 
chemical potential for the particle number conservation. 

Further, we extend the quasi-spin squeezed state for the pairing model 
to the O(4) model. 
Recently, to investigate the shape coexistence phenomena of nucleus, 
the simplest $O(4)$ algebraic model with the pairing and the 
quadrapole interactions has been reanalyzed beyond the Hartree-Fock-Bogoliubov 
and the random phase approximations.\cite{NW98,KNMM03}
As an extension of the paring type quasi-spin squeezed state, we attempt 
to construct possible quasi-spin squeezed states for the $O(4)$ model.

This paper is organized as follows. In the next section, the exact treatment 
and the Hartree-Fock approximation of pairing model\cite{K61,LM65,EG} are 
recapitulated containing notation. 
In \S 3, the time-dependent variational approach with the quasi-spin squeezed 
state for the pairing model is formulated. Also, it is shown that 
the ground state energy is well reproduced analytically 
in this dynamical approach. 
Against the former section, in \S 4, a static treatment, that is, the 
variation for the Hamiltonian with the chemical potential for the particle 
number conservation is carried out, and the ground state energy is 
estimated numerically in the pairing model. 
In \S 5, the extensions of the pairing-type squeezed state to the one for 
the $O(4)$ model are given. The energy expectation value for the ground state 
is calculated numerically by using two candidates of the squeezed states 
for the $O(4)$ model. The expectation values for the various operators 
with respect to two squeezed states are summarized in Appendix A and B, 
respectively. 
The last section is devoted to a summary. 

\section{Recapitulation of exact solution and coherent state approximation 
for pairing model with single energy level}

In this section, the exact treatment of an exactly solvable 
quantum many-fermion model, which is called the pairing model, 
is reviewed for later convenience. 
Also, the coherent state approximation, which corresponds to 
the BCS approximation, is recapitulated. 

\subsection{Exact solution for the pairing model}

We investigate a simple many-fermion system in which there exists 
$N$ identical fermions in a single spherical orbit with pairing interaction. 
The single particle state is specified by a set of quantum number 
$(j,m)$, where $j$ and $m$ represent the magnitude of angular momentum of 
the single particle state and 
its projection to the $z$-axis, respectively. 
Thus, let us start with the following Hamiltonian : 
\begin{equation}\label{2-1}
{\hat H}=\epsilon\sum_m{\hat c}^\dagger_{m}{\hat c}_m-\frac{G}{4}
\sum_{m}(-)^{j-m}{\hat c}_m^{\dagger}{\hat c}_{-m}^{\dagger}
\sum_{m}(-)^{j-m}{\hat c}_{-m}{\hat c}_m\ ,
\end{equation}
where $\epsilon$ and $G$ represent the single particle energy and 
the force strength, respectively. The operators ${\hat c}_m$ and 
${\hat c}_m^{\dagger}$ are the fermion annihilation and 
creation operators with the quantum number 
$m$, which obey the anti-commutation relations : 
\begin{equation}\label{2-2}
\{\ {\hat c}_m\ , \ {\hat c}_{m'}^{\dagger}\ \}=\delta_{mm'} \ , \qquad
\{\ {\hat c}_m\ , \ {\hat c}_{m'}\ \}=
\{\ {\hat c}_m^{\dagger}\ , \ {\hat c}_{m'}^{\dagger}\ \}=0 \ . 
\end{equation}
We introduce the following operators : 
\begin{equation}\label{2-3}
{\hat S}_+=\frac{1}{2}
\sum_{m}(-)^{j-m}{\hat c}_m^{\dagger}{\hat c}_{-m}^{\dagger} \ , \quad
{\hat S}_-=\frac{1}{2}
\sum_{m}(-)^{j-m}{\hat c}_{-m}{\hat c}_{m} \ , \quad
{\hat S}_0=\frac{1}{2}(\sum_m{\hat c}^\dagger_{m}{\hat c}_m-\Omega)\ , 
\end{equation}
where $\Omega$ represents the half of the degeneracy :$\Omega=j+1/2$. 
These operators compose the $su(2)$-algebra : 
\begin{equation}\label{2-4}
[{\hat S}_+ , {\hat S}_-]=2{\hat S}_0 \ , \qquad
[{\hat S}_0 , {\hat S}_{\pm}]=\pm{\hat S}_{\pm} \ .
\end{equation}
Thus, these operators are called the quasi-spin operators.\cite{K61,LM65} 
Then, the Hamiltonian (\ref{2-1}) can be rewritten in terms of 
the quasi-spin operators as 
\begin{eqnarray}\label{2-5}
{\hat H}&=&2\epsilon({\hat S}_0+S_j)-G{\hat S}_+{\hat S}_- \nonumber\\
&=&\epsilon {\hat N}-G{\hat S}_+{\hat S}_- \ , \\
S_j&=&\Omega/2\ (=(j+1/2)/2)\ , \nonumber
\end{eqnarray}
where ${\hat N}$ represents the number operator : 
\begin{equation}\label{2-6}
{\hat N}=\sum_{m}{\hat c}_m^{\dagger}{\hat c}_m\ .
\end{equation}

As is well known, the eigenstates and eigenvalues for this Hamiltonian 
are easily obtained. From $[{\hat S}_0 , {\hat H}]=0$, 
the eigenstates are given in terms of the eigenstates of 
${\hat S}_0$ : 
\begin{eqnarray}\label{2-7}
& &{\hat {\mib S}}^2\ket{S,S_0}=S(S+1)\ket{S,S_0} \ , \quad
{\hat S}_0\ket{S,S_0}=S_0\ket{S,S_0} \ , \nonumber\\
& &{\hat S}_{\pm}\ket{S,S_0}=\sqrt{(S\mp S_0)(S\pm S_0 +1)}\ket{S,S_0\pm 1} \ , 
\end{eqnarray}
where ${\hat {\mib S}}^2={\hat S}_0^2+({\hat S}_+{\hat S}_-+
{\hat S}_-{\hat S}_+)/2$. 
Thus, we obtain the eigenvalue equation 
\begin{equation}\label{2-8}
{\hat H}\ket{S,S_0}=[
2\epsilon (S_0+S_j)-G(S+S_0)(S-S_0+1)]\ket{S,S_0} 
\equiv E_\nu \ket{S,S_0}\ .
\end{equation}
Here, from the relation ${\hat S}_0={\hat N}/2-S_j$, 
the eigenvalue $S_0$ can be written in terms of particle number $N$ as 
$S_0=N/2-S_j$. Further, if we introduce the variable $\nu$ as 
\begin{equation}\label{2-9}
S=S_j-\nu/2 \ , \quad \nu=0,1,\cdots , 2S_j \ (=\Omega) \ ,
\end{equation}
the energy eigenvalue $E_\nu$ is given by 
\begin{equation}\label{2-10}
E_\nu=\epsilon N-\frac{1}{4}GN(1-\nu/N)(2\Omega-N-\nu+2) \ .
\end{equation}
Thus, the ground state energy can be obtained by setting $\nu=0$ as 
\begin{equation}\label{2-11}
E_0=\epsilon N-\frac{1}{4}GN\Omega
\left(2-\frac{N}{\Omega}+\frac{2}{\Omega}\right)\ . 
\end{equation}

\subsection{Coherent state approximation}

Next, we review the coherent state approach to this pairing model, 
which is identical with the BCS approximation to the pairing model 
consisting of the single energy level.

The $su(2)$-coherent state is given as 
\begin{eqnarray}\label{2-12}
& &\ket{\phi(\alpha)}=U(f)\ket{0} \ , \nonumber\\
& &U(f)=\exp (f{\hat S}_+-f^*{\hat S}_-) \ , \nonumber\\
& &\ket{0}=\ket{S=S_j, S_0=-S_j}\ , \qquad ({\hat S}_-\ket{0}=0)\ .
\end{eqnarray}

We impose the canonicity condition :
\begin{equation}\label{2-13}
\bra{\phi(\alpha)} \frac{\partial}{\partial \xi} \ket{\phi(\alpha)}
=\frac{1}{2}\xi^* \ , \qquad
\bra{\phi(\alpha)} \frac{\partial}{\partial \xi^*} \ket{\phi(\alpha)}
=-\frac{1}{2}\xi \ . 
\end{equation}
A possible solution of the above canonicity condition is presented as 
\begin{equation}\label{2-14}
\xi=\sqrt{2S_j}\frac{f}{|f|}\sin |f| \ , \qquad
\xi^*=\sqrt{2S_j}\frac{f^*}{|f|}\sin |f| \ .
\end{equation}
Then, the expectation values are obtained as 
\begin{eqnarray}
& &\bra{\phi(\alpha)}{\hat S}_+\ket{\phi(\alpha)}
=\xi^*\sqrt{2S_j-|\xi|^2} \ , \nonumber\\
& &\bra{\phi(\alpha)}{\hat S}_-\ket{\phi(\alpha)}
=\sqrt{2S_j-|\xi|^2}\ \xi \ , \nonumber\\
& &\bra{\phi(\alpha)}{\hat S}_0\ket{\phi(\alpha)}
=|\xi|^2-S_j \ , \label{2-15}\\
& &\bra{\phi(\alpha)}{\hat S}_+{\hat S}_-\ket{\phi(\alpha)}
=2S_j|\xi|^2-\left(1-\frac{1}{2S_j}\right)|\xi|^4 \ , 
\label{2-16}
\end{eqnarray}
Thus, the expectation value of Hamiltonian ${\hat H}$ 
and number operator ${\hat N}$ are given as 
\begin{eqnarray}\label{2-17}
& &\bra{\phi(\alpha)}{\hat H}\ket{\phi(\alpha)}
=\epsilon \cdot 2|\xi|^2-G\left(2S_j|\xi|^2-
\left(1-\frac{1}{2S_j}\right)|\xi|^4\right)
\equiv E \ , \nonumber\\
& &\bra{\phi(\alpha)}{\hat N}\ket{\phi(\alpha)}=2|\xi|^2 \equiv N \ .
\end{eqnarray}
If total particle number conserves, that is, $N=$constant, then, 
the energy expectation value $E$ is obtained as a function of $N$ : 
\begin{equation}\label{2-18}
E=\epsilon N-\frac{1}{4}GN\Omega\left(2-\frac{N}{\Omega}
+\frac{N}{\Omega^2}\right) \ .
\end{equation}

It is interesting to compare the exact ground state energy (\ref{2-11}) 
and the BCS approximated energy (\ref{2-18}). The last term of (\ref{2-18}) 
is different from that of the exact result in (\ref{2-11}). 
Let us assume that the particle number $N$ and the half of the 
degeneracy $\Omega$ are the same order of magnitude. 
If $\Omega$ (or $N$) is large, both the last terms of the exact eigenvalue 
(\ref{2-11}) and the approximate ground state energy (\ref{2-18}) 
can be neglected. Thus, the coherent state approximation presents 
good result for the ground state energy. This situation is similar 
to large $N$ limit which occurs in the several field of physics. 

In the next section, we try to reproduce the last term in a certain 
approximate approach, namely, the time-dependent variational 
approach with quasi-spin squeezed state.

\section{Dynamical approach of quasi-spin squeezed state for pairing model}

In this section it is shown that the ground state energy of 
a many-fermion system with pairing interaction can be well approximated 
analytically by using the time-dependent variational approach with 
a quasi-spin squeezed state. 

\subsection{Quasi-spin squeezed state}

In this subsection, the quasi-spin squeezed state is introduced following to 
Ref.\citen{TKY94}. 
First, we introduce the following operators :
\begin{equation}\label{3-1}
{\hat A}^{\dagger}=\frac{{\hat S}_+}{\sqrt{2S_j}} \ , \qquad
{\hat A}=\frac{{\hat S}_-}{\sqrt{2S_j}} \ , \qquad
{\hat N}=2({\hat S}_0+S_j) \ , 
\end{equation}
where ${\hat N}$ is identical with the number operator (\ref{2-6}). 
Then, the commutation relations can be expressed as 
\begin{equation}\label{3-2}
[{\hat A} , {\hat A}^{\dagger}]=1-\frac{{\hat N}}{2S_j} \ , \qquad
[{\hat N} , {\hat A}]=-2{\hat A} \ , \qquad
[{\hat N} , {\hat A}^{\dagger}]=2{\hat A}^{\dagger} \ . 
\end{equation}
Using the boson-like operator ${\hat A}^{\dagger}$, 
the $su(2)$-coherent state in (\ref{2-12}) can be recast into 
\begin{eqnarray}\label{3-3}
& &\ket{\phi(\alpha)}
=\frac{1}{\sqrt{\Phi(\alpha^*\alpha)}}
\exp(\alpha{\hat A}^{\dagger})\ket{0} \ , \nonumber\\
& &\Phi(\alpha^*\alpha)=\left(1+\frac{\alpha^*\alpha}{2S_j}\right)^{2S_j} \ , 
\end{eqnarray}
where $\alpha$ is related to $f$ in (\ref{2-12}) as 
$\alpha=\sqrt{2S_j}\cdot (f/|f|)\cdot \tan |f|$. 
The state $\ket{\phi(\alpha)}$ in (\ref{3-3}) is a vacuum state 
for the Bogoliubov transformed operator ${\hat a}_m$ :
\begin{equation}\label{3-4}
\hat{a}_m=U\hat{c}_m-V(-)^{j-m}\hat{c}^{\dagger}_{-m} \ , \qquad \hat{a}_m\ket{\phi(\alpha)}=0 \ .
\end{equation}
The coefficients $U$ and $V$ are given as 
\begin{equation}\label{3-5}
U=\frac{1}{\sqrt{1+\alpha^*\alpha/(2S_j)}} \ , \quad
V=\frac{\alpha}{\sqrt{2S_j}}\frac{1}{\sqrt{1+\alpha^*\alpha/(2S_j)}} \ , \quad
U^2+|V|^2=1 \ . 
\end{equation}
Of course, ${\hat a}_m$ and ${\hat a}_m^{\dagger}$ are fermion annihilation 
and creation operators and the anti-commutation relations are satisfied. 
By using the above Bogoliubov transformed operators, we introduce 
the following operators : 
\begin{equation}\label{3-6}
{\hat B}^{\dagger}
=\frac{1}{\sqrt{8S_j}}
\sum_m (-)^{j-m}{\hat a}_m^{\dagger}{\hat a}_{-m}^{\dagger} \ , \quad
{\hat B}
=\frac{1}{\sqrt{8S_j}}
\sum_m (-)^{j-m}{\hat a}_{-m}{\hat a}_{m} \ , \quad
{\hat M}=\sum_m{\hat a}_m^{\dagger}{\hat a}_m \ .
\end{equation}
Then, the state $\ket{\phi(\alpha)}$ satisfies 
\begin{equation}\label{3-7}
{\hat B}\ket{\phi(\alpha)}={\hat M}\ket{\phi(\alpha)}=0 \ .
\end{equation}
Further, the commutation relations are as follows :
\begin{equation}\label{3-8}
[{\hat B} , {\hat B}^{\dagger}]=1-\frac{{\hat M}}{2S_j} \ , \qquad
[{\hat M} , {\hat B}]=-2{\hat B} \ , \qquad
[{\hat M} , {\hat B}^{\dagger}]=2{\hat B}^{\dagger} \ . 
\end{equation}
The quasi-spin squeezed state can be constructed on the $su(2)$-coherent 
state $\ket{\phi(\alpha)}$ by using the above 
boson-like operator ${\hat B}^{\dagger}$ as is similar to the ordinary 
boson squeezed state : 
\begin{eqnarray}\label{3-9}
& &\ket{\psi(\alpha,\beta)}
=\frac{1}{\sqrt{\Psi(\beta^*\beta)}}\exp \left(
\frac{1}{2}\beta{\hat B}^{\dagger 2}\right) \ket{\phi(\alpha)} \ , \nonumber\\
& &\Psi(\beta^*\beta)=1+\sum_{k=1}^{[S_j]}\frac{(2k-1)!!}{(2k)!!}
\prod_{p=1}^{2k-1}\left(1-\frac{p}{2S_j}\right)(|\beta|^2)^k \ .
\end{eqnarray}
We call the state $\ket{\psi(\alpha,\beta)}$ the quasi-spin squeezed state.

\subsection{Canonicity conditions and expectation values}

In this subsection, we calculate the expectation values for various operators. 
The expectation values can be expressed in terms of the canonical 
variables which are introduced through the canonicity conditions. 
The same results derived in this subsection are originally given 
in the Lipkin model in Ref.\citen{Y94} at the first time 
and these results were used in Ref.\citen{TA03} to analyze the effects 
of quantum fluctuations in the Lipkin model. 
The Lipkin model is a simple 
many-fermion model with two energy levels in which the particle-hole 
interaction is active.\cite{LMG65} 
This model has the same algebraic structure as 
the pairing model, that is, the Hamiltonian can be expressed 
by the quasi-spin $su(2)$-generators. Thus, in our case, the same 
calculation as that given in Ref.\citen{Y94} are carried out. 

For the quasi-spin squeezed state $\ket{\psi(\alpha,\beta)}$ in (\ref{3-9}), 
the following expression is useful : 
\begin{eqnarray}\label{3-10}
\bra{\psi(\alpha,\beta)}\partial_z \ket{\psi(\alpha,\beta)}
&=&\frac{\Psi'(\beta^*\beta)}{\Psi(\beta^*\beta)}\frac{1}{2}
(\beta^*\partial_z \beta - \beta \partial_z \beta^*) \nonumber\\
& &+\left(1-\frac{2\beta^*\beta}{S_j}
\frac{\Psi'(\beta^*\beta)}{\Psi(\beta^*\beta)}\right)
\cdot \frac{1}{1+\alpha^*\alpha/(2S_j)}\frac{1}{2}
(\alpha^* \partial_z \alpha - \alpha \partial_z \alpha^*)\ , \nonumber\\
\Psi'(u)&\equiv&\frac{\partial \Psi(u)}{\partial u} \ , 
\end{eqnarray}
where $\partial_z=\partial/\partial z$. 
The canonicity conditions are imposed in order to introduce the sets of 
canonical variables $(X, X^*)$ and $(Y, Y^*)$ as follows :
\begin{eqnarray}\label{3-11}
& &\bra{\psi(\alpha,\beta)}\partial_X \ket{\psi(\alpha,\beta)}
=\frac{1}{2}X^* \ , \qquad
\bra{\psi(\alpha,\beta)}\partial_{X^*} \ket{\psi(\alpha,\beta)}
=-\frac{1}{2}X \ , \nonumber\\
& &\bra{\psi(\alpha,\beta)}\partial_Y \ket{\psi(\alpha,\beta)}
=\frac{1}{2}Y^* \ , \qquad
\bra{\psi(\alpha,\beta)}\partial_{Y^*} \ket{\psi(\alpha,\beta)}
=-\frac{1}{2}Y \ . 
\end{eqnarray}
Possible solutions for $X$ and $Y$ are obtained as 
\begin{eqnarray}\label{3-12}
& &\alpha=\frac{X}{\sqrt{1-\frac{X^*X}{2S_j}-\frac{4Y^*Y}{2S_j}}} \ , \quad
\alpha^*=\frac{X^*}{\sqrt{1-\frac{X^*X}{2S_j}-\frac{4Y^*Y}{2S_j}}} \ ,
\nonumber\\
& &\beta=\frac{Y}{\sqrt{K(Y^*Y)}} \ , \qquad
\beta^*=\frac{Y^*}{\sqrt{K(Y^*Y)}} \ , 
\end{eqnarray}
where $K(Y^*Y)$ is introduced and satisfies the relation 
\begin{equation}\label{3-13}
K(Y^*Y)\Psi(Y^*Y/K)=\Psi'(Y^*Y/K) \ .
\end{equation}
The expectation values for ${\hat B}$, ${\hat B}^{\dagger}$, 
${\hat M}$ and the products of these operators are easily obtained 
and are expressed in terms of the canonical variables as 
follows : 
\begin{subequations}\label{3-14}
\begin{eqnarray}
& &\bra{\psi(\alpha,\beta)}{\hat B}\ket{\psi(\alpha,\beta)}
=\bra{\psi(\alpha,\beta)}{\hat B}^{\dagger}\ket{\psi(\alpha,\beta)}=0 \ ,
\nonumber\\
& &\bra{\psi(\alpha,\beta)}{\hat M}\ket{\psi(\alpha,\beta)}
=4Y^*Y \ , 
\label{3-14a}\\
& &\bra{\psi(\alpha,\beta)}{\hat B}^2\ket{\psi(\alpha,\beta)}
=2Y\sqrt{K(Y^*Y)} \ , 
\nonumber\\
& &
\bra{\psi(\alpha,\beta)}{\hat B}^{\dagger 2}\ket{\psi(\alpha,\beta)}
=2Y^*\sqrt{K(Y^*Y)} \ , 
\label{3-14b}\\
& &\bra{\psi(\alpha,\beta)}{\hat B}^{\dagger}{\hat B}\ket{\psi(\alpha,\beta)}
=2(1-1/(2S_j))Y^*Y-\frac{2}{S_j}(Y^*Y)^2\cdot L(Y^*Y) \ , \nonumber\\
& &\bra{\psi(\alpha,\beta)}{\hat B}{\hat B}^{\dagger}\ket{\psi(\alpha,\beta)}
=\bra{\psi(\alpha,\beta)}{\hat B}^{\dagger}{\hat B}\ket{\psi(\alpha,\beta)}
+(1-2Y^*Y/S_j) \ , 
\label{3-14c}\\
& &\bra{\psi(\alpha,\beta)}{\hat M}^2\ket{\psi(\alpha,\beta)}
=16Y^*Y(1+Y^*Y\cdot L(Y^*Y)) \ , 
\label{3-14d}
\end{eqnarray}
where $L(Y^*Y)$ is defined and satisfies 
\begin{equation}\label{3-14e}
K(Y^*Y)^2\cdot L(Y^*Y)=\frac{\Psi''(\beta^*\beta)}{\Psi(\beta^*\beta)} \ .
\end{equation}
\end{subequations}

By using the relations between the original variables $\alpha$ and $\beta$ 
and the canonical variables $X$ and $Y$, 
the coefficients of the Bogoliubov transformation (\ref{3-5}), 
$U$ and $V$, are expressed as 
\begin{equation}\label{3-15}
U=\frac{\sqrt{1-\frac{X^*X}{2S_j}-\frac{4Y^*Y}{2S_j}}}
{\sqrt{1-\frac{4Y^*Y}{2S_j}}} \ , \quad
V=\frac{X}{\sqrt{2S_j}}\frac{1}{\sqrt{1-\frac{4Y^*Y}{2S_j}}} \ .
\end{equation}
Then, the operators ${\hat A}$, ${\hat A}^{\dagger}$ and ${\hat N}$, 
which are related to the quasi-spin operators ${\hat S}_-$, ${\hat S}_+$ 
and ${\hat S}_0$, respectively, in (\ref{3-1}), can be 
expressed as 
\begin{eqnarray}\label{3-16}
& &{\hat A}=\sqrt{2S_j}UV\left(1-\frac{\hat M}{2S_j}\right)
-V^2{\hat B}^{\dagger}+U^2{\hat B} \ , \nonumber\\
& &{\hat A}^{\dagger}=\sqrt{2S_j}UV^*\left(1-\frac{\hat M}{2S_j}\right)
+U^2{\hat B}^{\dagger}-V^{*2}{\hat B} \ , \nonumber\\
& &{\hat N}=4S_jV^*V\left(1-\frac{\hat M}{2S_j}\right)+\sqrt{2S_j}U
(V{\hat B}^{\dagger} +V^*{\hat B})+{\hat M} \ . 
\end{eqnarray}
Thus, the expectation values for ${\hat A}$, ${\hat A}^{\dagger}$, 
${\hat N}$ and the products of these operators are easily obtained 
and are expressed in terms of the canonical variables as 
follows : 
\begin{subequations}\label{3-17}
\begin{eqnarray}
& &\bra{\psi(\alpha,\beta)}{\hat A}\ket{\psi(\alpha,\beta)}
=X^*\sqrt{1-\frac{X^*X}{2S_j}-\frac{4Y^*Y}{2S_j}} \ , \nonumber\\
& &\bra{\psi(\alpha,\beta)}{\hat A}^{\dagger}\ket{\psi(\alpha,\beta)}
=\sqrt{1-\frac{X^*X}{2S_j}-\frac{4Y^*Y}{2S_j}}\ X \ , 
\nonumber\\
& &\bra{\psi(\alpha,\beta)}{\hat N}\ket{\psi(\alpha,\beta)}
=2X^*X+4Y^*Y \ , 
\label{3-17a}\\
& &\bra{\psi(\alpha,\beta)}{\hat A}^2\ket{\psi(\alpha,\beta)}
=\bra{\psi(\alpha,\beta)}{\hat A}\ket{\psi(\alpha,\beta)}^2\left[1-
\frac{1}{2S_j-4Y^*Y}\right] 
\nonumber\\
& &\qquad\qquad\qquad\qquad\qquad\quad
+\frac{U^2V^{2}}{2S_j}\left[
\bra{\psi(\alpha,\beta)}{\hat M}^2\ket{\psi(\alpha,\beta)}
-\bra{\psi(\alpha,\beta)}{\hat M}\ket{\psi(\alpha,\beta)}^2\right] \nonumber\\
& &\qquad\qquad\qquad\qquad\qquad\quad
+V^4\bra{\psi(\alpha,\beta)}{\hat B}^{\dagger 2}\ket{\psi(\alpha,\beta)}
+U^4\bra{\psi(\alpha,\beta)}{\hat B}^{2}\ket{\psi(\alpha,\beta)} 
\nonumber\\
& &\qquad\qquad\qquad\qquad\qquad\quad
-2U^2V^2\bra{\psi(\alpha,\beta)}{\hat B}^{\dagger}{\hat B}
\ket{\psi(\alpha,\beta)} 
\ , 
\nonumber\\
& &\bra{\psi(\alpha,\beta)}{\hat A}^{\dagger 2}\ket{\psi(\alpha,\beta)}
=\bra{\psi(\alpha,\beta)}{\hat A}^{\dagger}\ket{\psi(\alpha,\beta)}^2
\left[1-
\frac{1}{2S_j-4Y^*Y}\right] 
\nonumber\\
& &\qquad\qquad\qquad\qquad\qquad\quad
+\frac{U^2V^{*2}}{2S_j}\left[
\bra{\psi(\alpha,\beta)}{\hat M}^2\ket{\psi(\alpha,\beta)}
-\bra{\psi(\alpha,\beta)}{\hat M}\ket{\psi(\alpha,\beta)}^2\right] \nonumber\\
& &\qquad\qquad\qquad\qquad\qquad\quad
+U^4\bra{\psi(\alpha,\beta)}{\hat B}^{\dagger 2}\ket{\psi(\alpha,\beta)}
+V^{*4}\bra{\psi(\alpha,\beta)}{\hat B}^{2}\ket{\psi(\alpha,\beta)} 
\nonumber\\
& &\qquad\qquad\qquad\qquad\qquad\quad
-2U^2V^{*2}\bra{\psi(\alpha,\beta)}{\hat B}^{\dagger}{\hat B}
\ket{\psi(\alpha,\beta)} 
\ , 
\label{3-17b}\\
& &\bra{\psi(\alpha,\beta)}{\hat A}^{\dagger}{\hat A}\ket{\psi(\alpha,\beta)}
=\bra{\psi(\alpha,\beta)}{\hat A}^{\dagger}\ket{\psi(\alpha,\beta)}
\bra{\psi(\alpha,\beta)}{\hat A}\ket{\psi(\alpha,\beta)}
\nonumber\\
& &\qquad\qquad\qquad\qquad\qquad\quad
+\frac{(X^*X)^2}{2S_j(2S_j-4Y^*Y)}
\nonumber\\
& &\qquad\qquad\qquad\qquad\qquad\quad
+\frac{U^2V^*V}{2S_j}
\left[
\bra{\psi(\alpha,\beta)}{\hat M}^2\ket{\psi(\alpha,\beta)}
-\bra{\psi(\alpha,\beta)}{\hat M}\ket{\psi(\alpha,\beta)}^2\right]
\nonumber\\
& &\qquad\qquad\qquad\qquad\qquad\quad
-U^2V^2\bra{\psi(\alpha,\beta)}{\hat B}^{\dagger 2}\ket{\psi(\alpha,\beta)}
-U^2V^{*2}\bra{\psi(\alpha,\beta)}{\hat B}^{2}\ket{\psi(\alpha,\beta)}
\nonumber\\
& &\qquad\qquad\qquad\qquad\qquad\quad
+(1-2U^2V^*V)
\bra{\psi(\alpha,\beta)}{\hat B}^{\dagger}{\hat B}\ket{\psi(\alpha,\beta)}\ ,
\nonumber\\
& &\bra{\psi(\alpha,\beta)}{\hat A}{\hat A}^{\dagger}\ket{\psi(\alpha,\beta)}
=\bra{\psi(\alpha,\beta)}{\hat A}^{\dagger}{\hat A}\ket{\psi(\alpha,\beta)}
+\left(1-\frac{2X^*X}{2S_j}-\frac{4Y^*Y}{2S_j}\right) \  .
\label{3-17c}
\end{eqnarray}
\end{subequations}
From (\ref{3-1}), the expectation values of quasi-spin operators 
are derived from (\ref{3-17a}) : 
\begin{eqnarray}\label{3-18}
& &\bra{\psi(\alpha,\beta)}{\hat S}_+\ket{\psi(\alpha,\beta)}
=X^*\sqrt{2S_j-X^*X-4Y^*Y} \ , \nonumber\\
& &\bra{\psi(\alpha,\beta)}{\hat S}_-\ket{\psi(\alpha,\beta)}
=\sqrt{2S_j-X^*X-4Y^*Y}\ X \ , \nonumber\\
& &\bra{\psi(\alpha,\beta)}{\hat S}_0\ket{\psi(\alpha,\beta)}
=X^*X+2Y^*Y-S_j \ . 
\end{eqnarray}
By comparing with (\ref{2-15}), the variable $2|Y|^2$ represents 
the quantum fluctuations.

\subsection{Time-dependent variational approach with quasi-spin 
squeezed state for pairing model}

The model Hamiltonian (\ref{2-5}) can be expressed in terms of 
the fermion number operator ${\hat N}$ and the boson-like operators 
${\hat A}$ and ${\hat A}^{\dagger}$ as 
\begin{equation}\label{3-19}
{\hat H}=\epsilon {\hat N}-2S_j G{\hat A}^{\dagger}{\hat A} \ .
\end{equation}
Thus, the expectation value of this Hamiltonian is easily 
obtained by using (\ref{3-17}). 
We denote it as $H_{\rm sq}$ : 
\begin{equation}\label{3-20}
H_{\rm sq}=\bra{\psi(\alpha,\beta)}{\hat H}\ket{\psi(\alpha,\beta)} \ .
\end{equation}
The dynamics of this system can be investigated approximately by deriving 
the time-dependence of the canonical variables $(X, X^*)$ and $(Y, Y^*)$. 
The expectation value for the time-derivative is calculated as 
\begin{equation}\label{3-21}
\bra{\psi(\alpha,\beta)}i\partial_t\ket{\psi(\alpha,\beta)}
=\frac{i}{2}(X^*{\dot X}-{\dot X}^* X+Y^*{\dot Y}-{\dot Y}^* Y) \ ,
\end{equation}
where ${\dot X}$ represents the time-derivative of $X$ and so on. 
The time-dependence of $X$, $X^*$, $Y$ and $Y^*$ is derived from the 
time-dependent variational principle : 
\begin{equation}\label{3-22}
\delta\int \bra{\psi(\alpha,\beta)}i\partial_t-{\hat H}
\ket{\psi(\alpha,\beta)}dt=0 \ .
\end{equation}

Hereafter, we assume that $|Y|^2 \ll 1$ because $|Y|^2$ means the quantum 
fluctuations. 
Then, $K(Y^*Y)$ and $L(Y^*Y)$ defined in (\ref{3-13}) and (\ref{3-14e}), 
respectively, can be evaluated by the expansion with respect to $|Y|^2$. 
As a result, we obtain 
\begin{eqnarray}\label{3-23}
K(Y^*Y)&=&\frac{1}{2}\left(1-\frac{1}{2S_j}\right)
+\left(1-\frac{7}{2S_j}+\frac{9}{(2S_j)^2}\right)Y^*Y+\cdots \ . 
\nonumber\\
L(Y^*Y)&=&\frac{1}{1-\frac{1}{2S_j}}3\left(1-\frac{2}{2S_j}\right)\!\!
\left(1-\frac{3}{2S_j}\right) \nonumber\\
& &-\frac{48}{2S_j}\frac{1}{(1-\frac{1}{2S_j})^2}
\left(1-\frac{2}{2S_j}\right)\!\!\left(1-\frac{3}{2S_j}\right)\!\!
\left(1-\frac{4}{2S_j}\right)Y^*Y+\cdots \ . \qquad
\end{eqnarray}

Further, we introduce the action-angle variables instead of 
$(X,X^*)$ and $(Y, Y^*)$ as 
\begin{eqnarray}\label{3-24}
& &X=\sqrt{n_X}e^{-i\theta_X} \ , X^*=\sqrt{n_X}e^{i\theta_X} \ , 
\nonumber\\
& &Y=\sqrt{n_Y}e^{-i\theta_Y} \ , Y^*=\sqrt{n_Y}e^{i\theta_Y} \ . 
\end{eqnarray}
Then, the expectation values for the Hamiltonian, the time-derivative and 
the number operator can be expressed as 
\begin{subequations}\label{3-25}
\begin{eqnarray}
& &H_{\rm sq}=\epsilon(2n_X+4n_Y) -G\biggl[2S_j n_X-n_X^2
+\frac{n_X^2}{2S_j} \nonumber\\
& &\qquad\quad
  -2\sqrt{2}\sqrt{1-\frac{1}{2S_j}}\left(1-\frac{n_X}{2S_j}\right)
n_X\sqrt{n_Y}\cos (2\theta_X-\theta_Y) \nonumber\\
& &\qquad\quad
  +2\left(2S_j-1-4n_X+\frac{10}{2S_j}n_X
  +\frac{2}{2S_j}n_X^2-\frac{8}{(2S_j)^2}n_X^2\right)n_Y 
+{\rm O}(n_Y^{3/2}) \biggl] 
\ , \nonumber\\
& &\label{3-25a}\\
& &\bra{\psi(\alpha,\beta)}i\partial_t\ket{\psi(\alpha,\beta)}
=(n_X{\dot \theta}_X+n_Y{\dot \theta}_Y) \ , 
\label{3-25b}\\
& &N=\bra{\psi(\alpha,\beta)}{\hat N}\ket{\psi(\alpha,\beta)}
=2n_X+4n_Y \ . 
\label{3-25c}
\end{eqnarray}
\end{subequations}

From the time-dependent variational principle (\ref{3-22}), the following 
equations of motion are derived under the above-mentioned approximation : 
\begin{subequations}\label{3-26}
\begin{eqnarray}
& &
{\dot \theta}_X=\frac{\partial H_{\rm sq}}{\partial n_X}
\approx 2\epsilon-G\!\cdot\! 2\biggl[
S_j\!-\!n_X\!+\!\frac{n_X}{2S_j}\!-\!\sqrt{2}\sqrt{1-\frac{1}{2S_j}}
\!\!\left(
1-\frac{n_X}{S_j}\right)\!\!\sqrt{n_Y}\cos (2\theta_X-\theta_Y) \nonumber\\
& &\qquad\qquad\qquad\qquad\qquad
+4\left(-1+\frac{5}{4S_j}+\frac{1}{2S_j}n_X-\frac{1}{S_j^2}n_X\right)
n_Y\biggl] \ , 
\label{3-26a}\\
& &{\dot n}_X=-\frac{\partial H_{\rm sq}}{\partial \theta_X}
\approx -G\cdot 4\sqrt{2}\sqrt{1-\frac{1}{2S_j}}\!\left(1-\frac{n_X}{2S_j}
\right)\!n_X\sqrt{n_Y}\sin (2\theta_X-\theta_Y) \ , 
\label{3-26b}\\
& &
{\dot \theta}_Y=\frac{\partial H_{\rm sq}}{\partial n_Y}
\approx 4\epsilon-G\biggl[
-\sqrt{2}\sqrt{1-\frac{1}{2S_j}}\!\!\left(
1-\frac{n_X}{2S_j}\right)\!\!\frac{n_X}{\sqrt{n_Y}}\cos (2\theta_X-\theta_Y) 
\nonumber\\
& &\qquad\qquad\qquad\qquad\qquad
+2\left(-1+2S_j-4n_X+\frac{5}{S_j}+\frac{1}{S_j}n_X^2-\frac{2}{S_j^2}n_X^2
\right)\biggl] \ , 
\label{3-26c}\\
& &{\dot n}_Y=-\frac{\partial H_{\rm sq}}{\partial \theta_Y}
\approx G\cdot 2\sqrt{2}\sqrt{1-\frac{1}{2S_j}}\!\left(1-\frac{n_X}{2S_j}
\right)\!n_X\sqrt{n_Y}\sin (2\theta_X-\theta_Y) \ . 
\label{3-26d}
\end{eqnarray}
\end{subequations}
It is found from (\ref{3-26b}) and (\ref{3-26d}) that the total 
fermion number $N$ in (\ref{3-25c}) is conserved, that is, 
\begin{equation}\label{3-27}
{\dot N}=2{\dot n}_X+4{\dot n}_Y=0 \ . 
\end{equation}

\subsection{Dynamical approach to the ground state energy}

It should be noted here that $n_X \leq N/2$ from (\ref{3-25c}) and 
$N \leq \Omega/2=S_j$. Thus, the inequality $1-n_X/2S_j \geq 0$ 
is obtained. From the approximated energy expectation value (\ref{3-25a}) for 
$G>0$, 
the energy minimum is then obtained in the case $\cos(2\theta_X-\theta_Y)=-1$, 
namely, 
\begin{equation}\label{3-28}
\theta_Y=2\theta_X+\pi \ . 
\end{equation}
Since the energy is minimal in the ground state, 
the relation (\ref{3-28}) should be satisfied at any time. 
In order to assure the above-mentioned situation, the following 
consistency condition should be obeyed : 
\begin{equation}\label{3-29}
{\dot \theta}_Y=2{\dot \theta}_X \ . 
\end{equation}
Thus, from the equations of motion (\ref{3-26a}) and (\ref{3-26c}), 
under the approximation of small $n_Y$, the consistency condition 
(\ref{3-29}) gives the following expression of $n_Y$ 
in the lowest order approximation of $n_Y$ : 
\begin{eqnarray}\label{3-30}
\sqrt{n_Y}&\approx& \frac{\sqrt{1-\frac{1}{2S_j}}}
{2\sqrt{2}(1-\frac{2}{S_j})}\cdot
\left[1+\frac{1}{2n_X}\frac{1}{(1-\frac{n_X}{2S_j})(1-\frac{2}{S_j})}
\right]^{-1} \ .
\end{eqnarray}
Thus, by substituting (\ref{3-28}) and 
$\sqrt{n_Y}$ in (\ref{3-30}) under the lowest order 
approximation of $n_Y$ into 
the energy expectation value (\ref{3-25a}), 
and by performing the approximation 
of large $N$ or large $\Omega (=2S_j)$ approximation, we obtain the 
ground state energy as 
\begin{equation}\label{3-31}
H_{\rm sq}=\epsilon N-\frac{1}{4}GN\Omega
\left(2-\frac{N}{\Omega}+\frac{2}{\Omega} + 
{\rm O}(1/N\Omega, 1/\Omega^2, 1/N^2)\right) \ . 
\end{equation}
This result reproduces the exact energy eigenvalue (\ref{2-18}) 
by neglecting the higher order term of $1/N\Omega$, $1/\Omega^2$ and 
$1/N^2$ for large $N$ and $\Omega$ limit. 
Thus, the quasi-spin squeezed state presents a good approximation 
in the time-dependent variational approach to the pairing model. 
In this approach, the existence of the rotational motion in the phase 
space consisting of $(n_X, \theta_X; n_Y, \theta_Y)$ plays the important 
role. The angle variables for rotational motion, $\theta_X$ and $\theta_Y$, 
are consistently changed in (\ref{3-29}). 
This consistency condition is essential to reproduce the exact 
energy for the ground state under the large $N$ and $\Omega$ limit. 
The approximation corresponds to so-called large $N$ approximation. 
In general, it is known that the large $N$ expansion at zero temperature 
corresponds to $\hbar$ expansion. In this sense, the time-dependent 
variational approach with the quasi-spin squeezed state 
gives the approximation including the higher order quantum fluctuations 
than $\hbar$ if any expansion is not applied.

\section{Numerical estimation of ground state energy 
using quasi-spin squeezed state}

In the previous section, it has been shown that the exact ground state 
energy for the pairing model can be well reproduced in the large 
$N$ and $\Omega$ approximation in the time-dependent variational 
approach with the quasi-spin squeezed state. 
In that treatment, the rotational motion, which is originated from the 
use of the number violating state, plays the essential role to reproduce 
the ground state energy. 
In this section, we evaluate the expectation value for the ground state 
energy numerically by using the quasi-spin squeezed state 
without the expansion of $N$ and $\Omega$. 
Instead the time-dependent variational approach with the consistency 
condition developed in the previous section, the fermion umber conservation 
is guaranteed by introducing the chemical potential. 

We impose the minimization condition for the following quantities : 
\begin{eqnarray}\label{4-1}
\delta\langle {\hat H}'\rangle &\equiv& 
\delta\langle {\hat H}-\mu{\hat N} \rangle \nonumber\\
&=&\langle (\epsilon-\mu){\hat N}-2S_j G{\hat A}^{\dagger}{\hat A} 
\rangle =0 \ , 
\end{eqnarray}
where $\mu$ represents the chemical potential and $\langle \cdots \rangle$ 
denotes the expectation value with respect to the quasi-spin squeezed state. 
The expectation values for ${\hat N}$ and ${\hat A}^{\dagger}{\hat A}$ 
have been already given in (\ref{3-17}). 
The variation can be carried out the variational parameters 
$(\alpha, \alpha^*)$ and $(\beta, \beta^*)$. 
If we put $\beta=\beta^*=0$, 
the state is reduced to the $su(2)$-coherent state.

In Fig.1, the ground state energy with the unit $\epsilon$ 
is depicted in the case $N=\Omega=8$. 
The horizontal axis represents the force strength $G$ of the pairing 
interaction. The dotted curve, dot-dashed and solid curves 
represent the exact energy eigenvalue, the expectation value of the 
Hamiltonian with respect to the coherent state 
and the quasi-spin squeezed, respectively. 
\begin{figure}[t]
  \epsfxsize=7cm  
  \centerline{\epsfbox{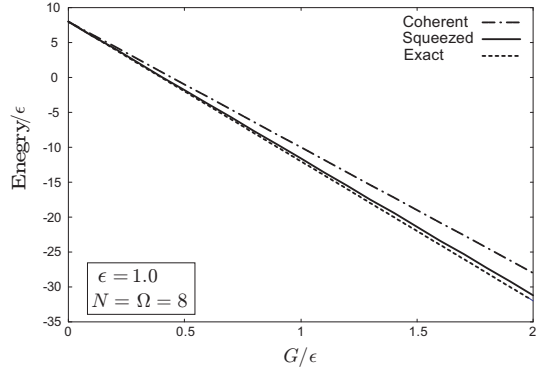}}
  \caption{Energy expectation values with respect to the coherent state 
  (dot-dashed curve) and the squeezed state (solid curve) 
  are depicted together with the exact eigenvalues (dotted curve) in the 
  case $N=\Omega=8$. 
  The horizontal axis represents $G$ with the unit $\epsilon$. }
   \label{fig:1}
\end{figure}
%
\begin{figure}[t]
  \epsfxsize=7cm  
  \centerline{\epsfbox{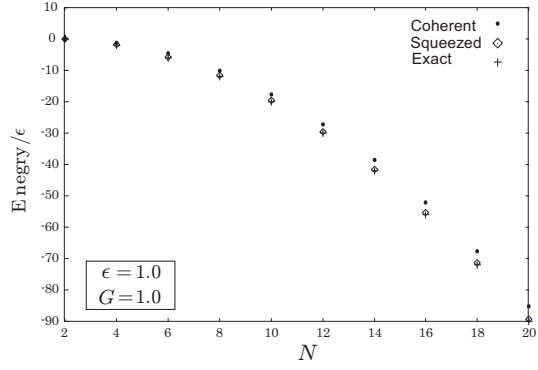}}
  \caption{Energy expectation values with respect to the coherent state 
  (dots) and the squeezed state (diamonds) 
  are depicted together with the exact eigenvalues (crosses) in the 
  case $\epsilon=1.0$ and $G=1.0$. 
  The horizontal axis represents $N$. }
   \label{fig:2}
\end{figure}
%
The result of the quasi-spin squeezed state well 
reproduces the exact eigenvalue for the wide range of $G$, comparing with 
the result by the coherent state.

In Fig.2, the energy is depicted in the case $\epsilon=1.0$ and $G=1$. 
The horizontal axis represents the particle number $N$ with $\Omega=N$. 
The result obtained by using the quasi-spin squeezed state 
is almost same as the exact eigenvalue. 
These figures show that the squeezed state approach 
presents a good approximation.

\section{Extension to O(4) model with pairing plus quadrapole interactions}

In the previous sections, \S\S 3 and 4, the $su(2)$-algebraic model 
with the pairing interaction has been investigated by using 
the quasi-spin squeezed state. It has been shown that 
the ground state energy has been well reproduced compared with the 
usual $su(2)$-coherent state. 
In this section, we try to extend our squeezed state approach to the 
$O(4)$-algebraic model with both the pairing and the quadrapole 
interactions in the many-fermion system such as nucleus.

\subsection{O(4) model with pairing and quadrapole interactions}

Let us start with the single-$j$ shell model, where $j$ represents 
the angular momentum quantum number. Thus, the degeneracy $2\Omega$ is 
$2\Omega=2j+1$. 
The pairing and the quadrapole interactions are active in this model. 
The Hamiltonian can be expressed as\cite{M82_} 
\begin{equation}\label{5-1}
{\hat H}_{\rm O(4)}=\epsilon {\hat N}-G{\hat P}^{\dagger}{\hat P}
-\frac{\chi}{2}{\hat Q}^2 \ . 
\end{equation}
Here, we define the following operators in terms of the 
fermion annihilation and creation operators 
$\{{\hat c}_m , {\hat c}_m^{\dagger} \}$ as 
\begin{eqnarray}\label{5-2}
& &{\hat P}^{\dagger}
=\sum_{m>0}^{j}{\hat c}_m^{\dagger}{\hat c}_{{\wtilde m}}^{\dagger}\ , \qquad
{\hat P}
=\sum_{m>0}^{j}{\hat c}_{{\wtilde m}}{\hat c}_{m}\ , \nonumber\\
& &{\hat Q}
=\sum_{m=-j}^{j}\sigma_{m}{\hat c}_m^{\dagger}{\hat c}_{m}\ , \qquad
{\hat N}
=\sum_{m=-j}^{j}{\hat c}_{m}^{\dagger}{\hat c}_{m}\ , \nonumber\\
& &{{\widetilde P}}^{\dagger}
=\sum_{m>0}^{j}\sigma_m 
{\hat c}_m^{\dagger}{\hat c}_{{\wtilde m}}^{\dagger}\ , \qquad
{{\widetilde P}}
=\sum_{m>0}^{j}\sigma_m
{\hat c}_{{\wtilde m}}{\hat c}_{m}\ , 
\end{eqnarray}
where 
\begin{equation}\label{5-3}
{\hat c}_{{\wtilde m}}=(-)^{j-m}{c_{-m}} \ , \qquad
\sigma_m=\cases{1\ , \quad {\rm for}\ |m|\leq \Omega/2 \cr
-1 \ , \quad {\rm for}\ |m|>\Omega/2 }
\end{equation}
and $\Omega$ represents the half of the degeneracy. In (\ref{5-2}), 
we also define ${\widetilde P}^{\dagger}$ and ${\widetilde P}$ for 
the later convenience. 

The Hamiltonian (\ref{5-1}) has the $O(4)$-algebraic structure. 
We can construct two $su(2)$-generators from the operators in (\ref{5-2}) : 
\begin{subequations}\label{5-4}
\begin{eqnarray}
& &{\hat S}_+^{\rm I}=({\hat P}^{\dagger}+{\wtilde P}^{\dagger})
=\sum_{0<m\leq \Omega/2} {\hat c}_m^{\dagger}{\hat c}_{\wtilde m}^{\dagger}
=({\hat S}_-^{{\rm I}})^{\dagger} \ , \nonumber\\
& &{\hat S}_0^{\rm I}=({\hat N}+{\hat Q}-\Omega)/4
=\left(2\sum_{|m|\leq\Omega/2}{\hat c}_m^{\dagger}{\hat c}_{\wtilde m}
-\Omega\right)/4 \ , 
\label{5-4a}\\
& &{\hat S}_+^{{\rm O}}=({\hat P}^{\dagger}-{\wtilde P}^{\dagger})
=\sum_{\Omega/2<m\leq \Omega} 
{\hat c}_m^{\dagger}{\hat c}_{\wtilde m}^{\dagger}
=({\hat S}_-^{{\rm O}})^{\dagger} \ , \nonumber\\
& &{\hat S}_0^{\rm O}=({\hat N}-{\hat Q}-\Omega)/4
=\left(2\sum_{\Omega/2<|m|\leq\Omega}{\hat c}_m^{\dagger}{\hat c}_{\wtilde m}
-\Omega\right)/4 \ , 
\label{5-4b}
\end{eqnarray}
\end{subequations}
where these operators satisfy the following commutation relations : 
\begin{eqnarray}\label{5-5}
& &[{\hat S}_+^{\rm I} , {\hat S}_-^{\rm I}]=2{\hat S}_0^{\rm I} \ , \qquad
[{\hat S}_0^{\rm I} , {\hat S}_{\pm}^{\rm I}]=\pm{\hat S}_{\pm}^{\rm I} \ , 
\nonumber\\
& &[{\hat S}_+^{\rm O} , {\hat S}_-^{\rm O}]=2{\hat S}_0^{\rm O} \ , \qquad
[{\hat S}_0^{\rm O} , {\hat S}_{\pm}^{\rm O}]=\pm{\hat S}_{\pm}^{\rm O} \ , 
\nonumber\\
& &[{\hat S}_{\pm,0}^{\rm I} , {\hat S}_{\pm,0}^{\rm O}]=0 \ . 
\end{eqnarray}
Thus, the sets $\{{\hat S}_{\pm,0}^{\rm I}\}$ and 
$\{{\hat S}_{\pm,0}^{\rm O}\}$
 give two sets of independent $su(2)$-generators. 
By expressing the operators ${\hat P}^{\dagger}$, ${\hat P}$ and 
${\hat N}$ inversely in terms of the above two $su(2)$-generators, 
the Hamiltonian in (\ref{5-1}) can be expressed in terms of two 
independent $su(2)$-generators. Thus, this model given by the Hamiltonian 
in (\ref{5-1}) is called $O(4)\ (\simeq su(2)\times su(2))$ model. 
This model can be solved exactly because the model space is spanned 
by two quasi-spin $su(2)$ states and the diagonalization is easily performed. 
Thus, the validity of an approximation can be checked.

An approach by the $O(4)$-coherent state corresponds to the 
Hartree-Fock approximation. The $O(4)$-coherent state 
can be constructed by the direct product of the two $su(2)$-coherent 
state as 
\begin{eqnarray}\label{5-6}
& &\ket{c}_{O(4)}=\ket{\alpha_I}\otimes \ket{\alpha_O}
\equiv \ket{\alpha_I,\alpha_O} \ , \nonumber\\
& &\quad \ket{\alpha_I}=\frac{1}{\sqrt{\Phi_I}}
\exp \left(\alpha_I\frac{{\hat S}_+^{\rm I}}{\sqrt{2S_I}}\right) \ket{0}_I
\ , \quad
\ket{\alpha_O}=\frac{1}{\sqrt{\Phi_O}}
\exp \left(\alpha_O\frac{{\hat S}_+^{\rm O}}{\sqrt{2S_O}}\right) \ket{0}_O
\ , \nonumber\\
& &\qquad\Phi_I=\left(1+\frac{\alpha_I^*\alpha_I}{2S_I}\right)^{2S_I} \ , 
\qquad
\Phi_O=\left(1+\frac{\alpha_O^*\alpha_O}{2S_O}\right)^{2S_O} \ , 
\nonumber\\
& &\qquad\quad
S_I=\Omega/4 \ , \qquad\qquad S_O=\Omega/4 \ . 
\end{eqnarray}
Thus, the variational parameters are $\alpha_I$, $\alpha_I^*$, $\alpha_O$ 
and $\alpha_O^*$ in $O(4)$-coherent state. 

For the number conservation, we introduce the chemical potential. 
Then, the variation with respect to $\alpha_I$, $\alpha_I^*$, $\alpha_O$ 
and $\alpha_O^*$ is carried out as  
\begin{equation}\label{5-7}
\delta \bra{\alpha_I,\alpha_O}{\hat H}_{O(4)}-\mu{\hat N} 
\ket{\alpha_I,\alpha_O}=0\ . 
\end{equation}
Figure 3 shows the exact ground state energy eigenvalues (dotted curve) and 
the energy expectation value calculated by the $O(4)$-coherent state 
(dashed curve). 
The model parameters $G$ and $\chi$ are parameterized as 
$G=\cos \theta$ and $2\chi= \sin \theta$, respectively,\cite{M82_} and 
$N=\Omega=8$ and $\epsilon =0$. 
It is found that, if the quadrapole interaction is dominant, that is, 
$\theta$ is large, the coherent 
state approximation presents good results for the ground state energy. 
However, the pairing interaction is dominant, 
the coherent state approximation is rather bad. 
There is a room to devise the approximate state in the $O(4)$ model
with the pairing and quadrapole interactions such as nucleus.

\subsection{Direct product of two quasi-spin squeezed states}

As is similar to the $su(2)$-squeezed state given in \S 3, let us 
construct the squeezed state for the $O(4)$ model. 
First, let us introduce the Bogoliubov transformed fermion annihilation 
and creation operators $({\hat a}_m, {\hat b}_m ; {\hat a}_m^{\dagger}, 
{\hat b}_m^{\dagger})$ : 
\begin{eqnarray}\label{5-8}
& &{\hat a}_m=U_I{\hat c}_m-V_I{\hat c}_{\wtilde m}^{\dagger}\ , 
\quad {\rm for}\ \ |m| \leq \Omega/2 \ , \nonumber\\
& &{\hat b}_m=U_O{\hat c}_m-V_O{\hat c}_{\wtilde m}^{\dagger}\ , 
\quad {\rm for}\ \ \Omega/2 < |m| \leq \Omega \ , 
\end{eqnarray}
where the coefficients of the Bogoliubov transformation are given as 
\begin{eqnarray}\label{5-9}
& &U_I=\frac{1}{\sqrt{1+\alpha_I^*\alpha_I/(2S_I)}} \ , \quad
V_I=\frac{\alpha_I}{\sqrt{2S_I}}\frac{1}{\sqrt{1+\alpha_I^*\alpha_I/(2S_I)}} 
\ , \quad U_I^2+|V|_I^2=1 \ , \nonumber\\
& &U_O=\frac{1}{\sqrt{1+\alpha_O^*\alpha_O/(2S_O)}} \ , \quad
V_O=\frac{\alpha_O}{\sqrt{2S_O}}\frac{1}{\sqrt{1+\alpha_O^*\alpha_O/(2S_O)}} 
\ , \quad U_O^2+|V|_O^2=1 \ . \nonumber\\
& &
\end{eqnarray}
Then, the boson-like operators are introduced like (\ref{3-6}) : 
\begin{eqnarray}\label{5-10}
& &{\hat B}_I^{\dagger}
=\frac{1}{\sqrt{2S_I}}
\sum_{0<m\leq \Omega/2}{\hat a}_m^{\dagger}{\hat a}_{\wtilde m}^{\dagger} \ , 
\>~\!
{\hat B}_I
=\frac{1}{\sqrt{2S_I}}
\sum_{0<m\leq \Omega/2} {\hat a}_{\wtilde m}{\hat a}_{m} \ ,\>
~\!
{\hat M}_I=\sum_{|m|\leq \Omega/2}{\hat a}_m^{\dagger}{\hat a}_m \ , 
\nonumber\\
& &{\hat B}_O^{\dagger}
=\frac{1}{\sqrt{2S_O}}
\sum_{\Omega/2<m\leq \Omega}
{\hat b}_m^{\dagger}{\hat b}_{\wtilde m}^{\dagger} \ , 
\>~\!
{\hat B}_O
=\frac{1}{\sqrt{2S_O}}
\sum_{\Omega/2<m\leq \Omega} {\hat b}_{\wtilde m}{\hat b}_{m} \ , 
\>~\! 
{\hat M}_O=\sum_{\Omega/2 <|m|}{\hat b}_m^{\dagger}{\hat b}_m \ . \nonumber\\
& &
\end{eqnarray}
Then, the following commutation relations are satisfied : 
\begin{eqnarray}\label{5-11}
& &[{\hat B}_I , {\hat B}_I^{\dagger}]=1-\frac{{\hat M}_I}{2S_I} \ , \qquad
[{\hat M}_I , {\hat B}_I]=-2{\hat B}_I \ , \qquad
[{\hat M}_I , {\hat B}_I^{\dagger}]=2{\hat B}_I^{\dagger} \ , \nonumber\\
& &[{\hat B}_O , {\hat B}_O^{\dagger}]=1-\frac{{\hat M}_O}{2S_O} \ , \quad
[{\hat M}_O , {\hat B}_O]=-2{\hat B}_O \ , \quad
[{\hat M}_O , {\hat B}_O^{\dagger}]=2{\hat B}_O^{\dagger} \ .  
\end{eqnarray}

The quasi-spin squeezed state $\ket{\psi}$ for $O(4)$ model may be constructed 
by the direct product of the two quasi-spin squeezed states, 
$\ket{\psi(\alpha_I,\beta_I)}$ and $\ket{\psi(\alpha_O,\beta_O)}$, 
which are defined as is similar to (\ref{3-9}) : 
\begin{eqnarray}\label{5-12}
& &\ket{\psi}=\ket{\psi(\alpha_I,\beta_I)}\otimes
\ket{\psi(\alpha_O,\beta_O)} \ , \nonumber\\
& &\ \ \ket{\psi(\alpha_I,\beta_I)}
=\frac{1}{\sqrt{\Psi(\beta_I^*\beta_I)}}
\exp\left(\frac{1}{2}\beta_I{\hat B}_I^{\dagger 2}\right)\ket{\alpha_I} 
\ , \nonumber\\
& &\ \ \ket{\psi(\alpha_O,\beta_O)}
=\frac{1}{\sqrt{\Psi(\beta_O^*\beta_O)}}
\exp\left(\frac{1}{2}\beta_O{\hat B}_O^{\dagger 2}\right)\ket{\alpha_O} \ , 
\\
& &\ \ \ \ \Psi(\beta_I\beta_I^*)=
1+\sum_{k=1}^{[S_I]}\frac{(2k-1)!!}{(2k)!!}(|\beta_I|^2)^k
\prod_{p=1}^{2k-1}\left(1-\frac{p}{2S_I}\right) \ ,  \nonumber\\
& &\ \ \ \ \Psi(\beta_O\beta_O^*)=
1+\sum_{k=1}^{[S_O]}\frac{(2k-1)!!}{(2k)!!}(|\beta_O|^2)^k
\prod_{p=1}^{2k-1}\left(1-\frac{p}{2S_O}\right) \ . \nonumber
\end{eqnarray}
Since the two $su(2)$-generators are expressed in terms of the 
above introduced boson-like operators in (\ref{5-10}) as 
\begin{eqnarray}\label{5-13}
& &{\hat S}_+^{\rm I}=2S_I\left(1-\frac{{\hat M}_I}{2S_I}\right)U_IV_I^*
+\sqrt{2S_I}(U_I^2{\hat B}_I^{\dagger}-V_I^{*2}{\hat B}_I) 
=({\hat S}_-^{\rm I})^{\dagger} \ , \nonumber\\
& &{\hat S}_0^{\rm I}=-S_I(U_I^2-|V_I|^2)\left(1-\frac{{\hat M}_I}{2S_I}
\right)+\sqrt{2S_I}U_I(V_I{\hat B}_I+V_I^*{\hat B}_I^{\dagger}) \ , \nonumber\\
& &{\hat S}_+^{\rm O}=2S_O\left(1-\frac{{\hat M}_O}{2S_O}\right)U_OV_O^*
+\sqrt{2S_O}(U_O^2{\hat B}_O^{\dagger}-V_O^{*2}{\hat B}_O) 
=({\hat S}_-^{\rm O})^{\dagger} \ , \nonumber\\
& &{\hat S}_0^{\rm O}=-S_O(U_O^2-|V_O|^2)\left(1-\frac{{\hat M}_O}{2S_O}
\right)+\sqrt{2S_O}U_O(V_O{\hat B}_O+V_O^*{\hat B}_O^{\dagger}) \ .
\end{eqnarray}
Also, the pairing operator ${\hat P}$, the quadrapole operator 
${\hat Q}$ and the number operator ${\hat N}$ can be expressed 
in terms of the above two sets of the $su(2)$-generators as 
\begin{equation}\label{5-14}
{\hat P}={\hat S}_-^{\rm I}+{\hat S}_-^{\rm O} \ , \quad
{\hat Q}=2({\hat S}_0^{\rm I}-{\hat S}_0^{\rm O}) \ , \quad
{\hat N}=2({\hat S}_0^{\rm I}+{\hat S}_0^{\rm O})+\Omega \ . 
\end{equation}
Thus, the expectation value of the Hamiltonian (\ref{5-1}) can be 
expressed by the $su(2)$-generators.

The variation are carried out with respect to the eight variational parameters 
$\alpha_{I,O}$, $\alpha_{I,O}^*$, $\beta_{I,O}$ and $\beta_{I,O}^*$ as 
\begin{equation}\label{5-15}
\delta \bra{\psi}{\hat H}_{O(4)}-\mu{\hat N} 
\ket{\psi}=0\ . 
\end{equation}
The expectation values are summarized in Appendix A. 
\begin{figure}[t]
  \epsfxsize=7cm  
  \centerline{\epsfbox{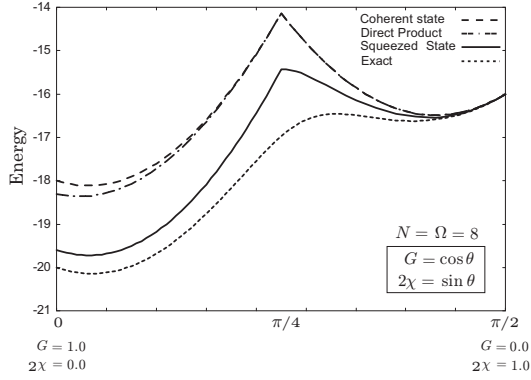}}
  \caption{Energy expectation values with respect to the coherent state 
  (dashed curve), the direct product of two quasi-spin squeezed state 
  (dot-dashed curve) and the fermionic squeezed state (solid curve) 
  are depicted together with the exact eigenvalues (dotted curve) in the 
  case $N=\Omega=8$. 
  The horizontal axis represents $\theta$ which parameterize the model 
  parameters $G=\cos \theta$ and $2\chi=\sin \theta$. }
   \label{fig:3}
\end{figure}
%
The energy expectation value for this model Hamiltonian with respect to 
the direct product of the two quasi-spin squeezed state are 
depicted in Fig. 3 (dot-dashed curve) 
compared with the exact eigenvalues (dotted curve) and the expectation 
values derived by using the $O(4)$-coherent state (dashed curve) 
in the case $N=\Omega=8$. 
The horizontal axis denotes $\theta$ where we parameterize the 
force strength $G$ and $\chi$ as 
$G=\cos \theta$ and $2\chi=\sin \theta$. 
It is found that the direct product of the two $su(2)$-spin squeezed states 
does not present good results for the ground state energy, especially in 
the region where the pairing interaction is dominant. 
The reason is as follows : If $\chi=0$, the $O(4)$ model is reduced to the 
pairing model investigated in \S\S 2$\sim$ 4. 
However, the direct product of the two quasi-spin squeezed state 
is not reduced to the state in (\ref{3-9}) because of the absence 
of the cross term ${\hat B}_I^{\dagger}{\hat B}_O^{\dagger}$, even if 
the parameters are set up as $\alpha_I=\alpha_O$ and $\beta_I=\beta_O$. 
Thus, in this case, the direct product 
does not include appropriate pairing correlations fully, and thus 
does not give a suitable squeezed state 
for many-fermion systems.

\subsection{Fermionic squeezed state}

In the previous subsection, the squeezed state for the $O(4)$ model 
has been constructed by the direct product of the two quasi-spin 
squeezed state. The extension of the $su(2)$-quasi-spin squeezed state 
to that for the $O(4)$ model may be natural from the viewpoint of the 
algebraic structure. 
However, the ground state energy is not reproduced so well. 
In this subsection, the trial state can be devised, 
which we call a fermionic squeezed state. 

We define another squeezed state $\ket{s}$ as 
\begin{eqnarray}\label{5-16}
& &\ket{s} =\frac{1}{\sqrt{\Gamma}}
{\rm exp}\left[ \beta_I \hat{B}_I^{\dagger}
+\beta_O \hat{B}_O^{\dagger}\right]^{2}\ket{c}_{O(4)} \ , 
\\
& &\ \Gamma=\sum_{n=0}^{[\Omega/2]}\frac{[(2n)!]^2}{(n!)^2}
\sum_{r=0}^{2n}\frac{|\beta_I|^{2r}|\beta_O|^{4n-2r}}{(2n-r)!r!}
\prod_{p=1}^{r-1}\left(1-\frac{p}{2S_I}\right)
\prod_{q=1}^{2n-r-1}\left(1-\frac{q}{2S_O}\right) \ . \nonumber
\end{eqnarray}
The state $\ket{s}$ can be reduced to the quasi-spin squeezed state 
if the conditions $\alpha_I=\alpha_O$ and $\beta_I=\beta_O$ are imposed. 
Thus, the pairing correlations are fully taken into account. 
The necessary expectation values for calculating the expectation value 
of the Hamiltonian are summarized in Appendix B. 
Thus, we can derive the variational equations in Eq.(\ref{5-15}). 
By solving the variational equations, we can obtain eight parameters 
$(\alpha_{I,O}, \alpha_{I,O}^*, \beta_{I,O}, \beta_{I,O}^*)$, and then 
the energy expectation value is evaluated. 
In Fig. 3, the energy expectation values calculated by using the state 
$\ket{s}$ are depicted (solid curve) 
compared with the energies of exact approach, 
coherent state approximation and the approximation by using the direct 
product of two quasi-spin squeezed state in (\ref{5-12}). 
It is shown that our squeezed state approach with the state $\ket{s}$ 
presents good results. 
\begin{figure}[t]
  \epsfxsize=7cm  
  \centerline{\epsfbox{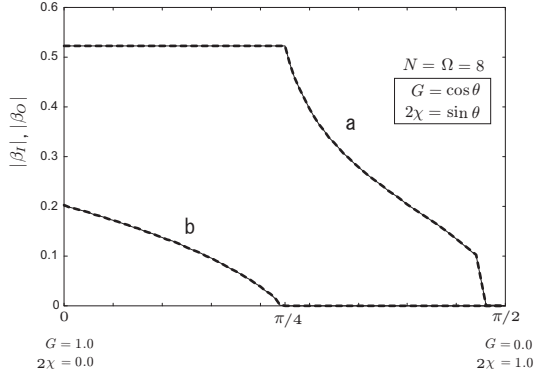}}
  \caption{The variables $|\beta_I|$ (solid curves) 
  and $|\beta_O|$ (dashed curves) are depicted in the 
  fermionic squeezed state approach (a) 
  and the approach with the direct product 
  of two quasi-spin squeezed state (b), respectively, 
  in the case $N=\Omega=8$. 
  The horizontal axis represents $\theta$ which parameterize the model 
  parameters $G=\cos \theta$ and $2\chi=\sin \theta$. }
   \label{fig:4}
\end{figure}
%
In Fig. 4, variables $|\beta_I|$ and $|\beta_O|$, which represent 
the quantum fluctuations and by which the particle-particle correlations 
are taken into account, are depicted in the case of the fermionic 
squeezed state (a) and the direct product of two quasi-spin squeezed state 
(b), respectively. 
In both cases, $|\beta_I|$ (solid curves) and $|\beta_O|$ (dashed curves) 
are almost the same. 
In all regions, the values $|\beta_I|$ and $|\beta_O|$ in the fermionic 
squeezed state approach are larger 
than those in the direct product of two quasi-spin squeezed state 
approach. Especially, in the fermionic squeezed state approach, 
it is found that the values are not negligible in the region where the 
pairing correlation is dominant. 
Thus, the pairing correlation is 
taken into account in this state. 
However, the values of $|\beta_I|$ and $|\beta_O|$ are rather small 
in the region where the quadrapole-quadrapole correlation is dominant. 
It may be concluded that the $O(4)$-coherent state is rather good state 
for describing the system in which the quadrapole correlation is rather 
strong.

\begin{figure}[t]
  \epsfxsize=7cm  
  \centerline{\epsfbox{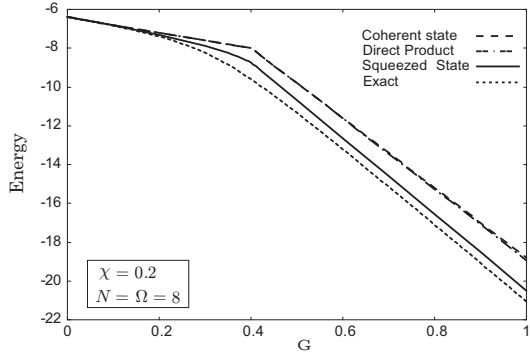}}
  \caption{Energy expectation values with respect to the coherent state 
  (dashed curve), the direct product of two quasi-spin squeezed state 
  (dot-dashed curve) and the fermionic squeezed state (solid curve) 
  are depicted together with the exact eigenvalues (dotted curve) in the 
  case $N=\Omega=8$ and $\chi=0.2$. 
  The horizontal axis represents $G$. }
   \label{fig:5}
\end{figure}
%
\begin{figure}[t]
  \epsfxsize=7cm  
  \centerline{\epsfbox{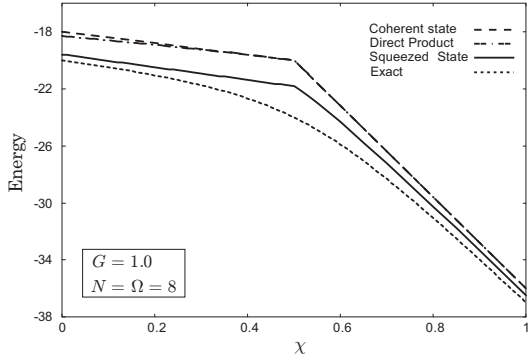}}
  \caption{Energy expectation values with respect to the coherent state 
  (dashed curve), the direct product of two quasi-spin squeezed state 
  (dot-dashed curve) and the fermionic squeezed state (solid curve) 
  are depicted together with the exact eigenvalues (dotted curve) in the 
  case $N=\Omega=8$ and $G=1$. 
  The horizontal axis represents $\chi$.}
   \label{fig:6}
\end{figure}
%
In Figs. 5 and 6, the energy expectation values with respect to 
the coherent state (dashed curve), the direct product of two 
quasi-spin squeezed state (dot-dashed curve) and the fermionic squeezed 
state (solid curve) are also depicted together with the exact eigenvalues 
(dotted curve) in the case $N=\Omega=8$ with $\chi=0.2$ (Fig.5) 
and $G=1.0$ (Fig.6), respectively, in which the horizontal 
axes represent $G$ (Fig.5) and $\chi$ (Fig.6). 
It is found that the fermionic squeezed state approach presents 
rather good results in all the parameter regions with $G$ and $\chi$.

\subsection{k-squeezed state}

In this subsection, we give one comment. 
Let us consider the $su(2)$ pairing model discussed in \S 2 $\sim$ \S 4. 
We can define the $k$-squeezed state as 
\begin{eqnarray}\label{6-1}
& &\ket{s}_k=\frac{1}{\sqrt{\Gamma_k}}\exp\left(\frac{w}{2}{\hat B}^{\dagger k}
\right)\ket{\phi(\alpha)} \ , \qquad (k \geq 3,\ ({\rm integer}))\ , 
\\
& &\Gamma_k=\sum_{n=0}^{[\Omega/k]}\frac{(kn)!}{4^n(n!)^2}|w|^{2n}
\prod_{p=1}^{kn-1}\left(1-\frac{p}{2S_j}\right) \ . \nonumber
\end{eqnarray}
In this state, the expectation values are calculated easily as 
\begin{eqnarray}\label{6-2}
& &{}_k\bra{s}{\hat B}\ket{s}_k={}_k\bra{s}{\hat B}^{\dagger}\ket{s}_k
={}_k\bra{s}{\hat B}^2\ket{s}_k={}_k\bra{s}{\hat B}^{\dagger 2}\ket{s}_k
=0 \ , \nonumber\\
& &{}_k\bra{s}{\hat M}\ket{s}_k=
\frac{1}{\Gamma_k}\sum_{n=0}^{[\Omega/k]}(2kn)\frac{(kn)!}{(n!)^2}
\frac{|w|^{2n}}{4^n}\prod_{p=1}^{kn-1}\left(1-\frac{p}{2S_j}\right) \ , 
\nonumber\\
& &{}_k\bra{s}{\hat M}^2\ket{s}_k=
\frac{1}{\Gamma_k}\sum_{n=0}^{[\Omega/k]}(2kn)^2\frac{(kn)!}{(n!)^2}
\frac{|w|^{2n}}{4^n}\prod_{p=1}^{kn-1}\left(1-\frac{p}{2S_j}\right) \ , 
\nonumber\\
& &{}_k\bra{s}{\hat B}{\hat B}^{\dagger}\ket{s}_k=
\frac{1}{\Gamma_k}\sum_{n=0}^{[\Omega/k]}\frac{(kn+1)!}{(n!)^2}
\frac{|w|^{2n}}{4^n}\prod_{p=1}^{kn}\left(1-\frac{p}{2S_j}\right) \ , 
\nonumber\\
& &{}_k\bra{s}{\hat B}^{\dagger}{\hat B}\ket{s}_k
={}_k\bra{s}{\hat B}{\hat B}^{\dagger}\ket{s}_k-
\left(1-\frac{{}_k\bra{s}{\hat M}\ket{s}_k}{2S_j}\right) \ , 
\end{eqnarray}
and the other combinations of the products of two operators give 
no finite values. 
In order to calculate the ground state energy, 
we impose the condition of minimization of the energy expectation value. 
As a result, it is concluded that $|w|=0$ is a solution for minimization. 
Thus, the $k$-squeezed state is reduced to the $su(2)$-coherent state 
in this pairing model. 
Therefore, the quasi-spin squeezed state gives a good state in order to 
take into account the pairing correlation.

\section{summary}
We have investigated a validity of the quasi-spin squeezed states as trial 
states in the variational method in the many-fermion systems with simple 
algebraic structures. First, we have applied the quasi-spin squeezed state to 
the pairing model by two approaches. One is dynamical approach, namely, the 
time-dependent variational approach. 
In this approach, the total fermion number is conserved automatically. 
We have imposed a consistency condition for the rotation in the phase 
space. As a result, the ground state energy is realized up to 
$O(1/\Omega^2, 1/N^2)$, which is better than the result of the 
coherent state approach, namely, the Hartree-Fock-Bogoliubov approximation. 
The other is to obtain the ground state energy directly in the 
energy minimum condition with the chemical potential. 
We have estimated the ground state energy numerically in the 
variational method, and compared the result with that of the coherent 
state approach and the exact energy eigenvalue. 
It has been shown that 
this quasi-spin squeezed state approach outrages the coherent state 
approach in all regions of the strength of the paring interaction.\\
\indent
Secondly, we have applied the quasi-spin squeezed state to the 
$O(4)$ model which has a $su(2)\times su(2)$ algebraic structure. 
Since we have the $su(2)$ quasi-spin squeezed state to take account 
the pairing correlation, we have adopted the direct product of two 
$su(2)$ quasi-spin squeezed states as a trial state. 
However, it has turned out that the quasi-spin squeezed state 
approach using the direct product of two quasi-spin squeezed states 
has not given good results compared with the $O(4)$-coherent state 
approach. 
The reason is that the pairing correlation can not be included 
appropriately, and the direct product of two quasi-spin squeezed states 
is not reduced to the quasi-spin squeezed state for pairing model 
when $\chi=0$. Thus, we have constructed another squeezed state free from 
the $O(4)$ algebra, which we call the fermionic squeezed state. 
This improved squeezed state includes partially the quadrapole correlation 
and is reduced to the quasi-spin squeezed state when two parameters are 
identical. 
Although the fermionic squeezed state has given rather good results 
than the direct product state, it may be not enough to include the 
quadrapole correlation when the quadrapole interaction is dominated 
($G<2\chi$). This is the further problem to find more suitable state. \\
\indent
For further application of a quasi-spin squeezed state approach,
it is interesting to investigate a nuclear $su(2)$-model which
interact with the environment represented by a harmonic 
oscillator.\cite{H01}
In this model, a certain case, a dissipative process
has been realized. However, in our previous treatment,\cite{H01}
 a quantum fluctuation has not been
took into account. 
Therefore, a quasi-spin squeezed state approach might be
suitable to introduce and investigate the effect of quantum fluctuation
into the system.

\acknowledgement 
One of the authors (Y.T.) would like to express his sincere thanks to 
Professor M. Yamamura who gives a chance to study about the time-dependent 
variational approach with the quasi-spin squeezed state.\cite{Y94} 
He also thanks Dr.~T.~Nakatsukasa for valuable discussion. 
He is partially supported by the Grants-in-Aid of the Scientific Research 
No. 15740156 from the Ministry of Education, Culture, Sports, Science and 
Technology in Japan.

\appendix
\section{Various expectation values by using the direct product of two
quasi-spin squeezed states}

In this appendix, we summarize the various expectation values with respect 
to the direct product of the quasi-spin squeezed states. 
We introduce new notations as 
$\Psi(\beta_I\beta_I^*)=\Psi_I$, 
$\Psi(\beta_O\beta_O^*)=\Psi_O$, 
$\Psi'_I=d\Psi_I/d|\beta_I|^2$ and  
$\Psi'_O=d\Psi_O/d|\beta_O|^2$. 

\def\<{\langle}     
\def\>{\rangle} 

\begin{eqnarray}
& &\< \psi|{\hat M}_{I}\hat{B}_I| \psi\>
=\<\psi|\hat{B}_I^{\dagger}{\hat M}_{I}| \psi\>
=\< \psi|\hat{B}_I{\hat M}_{I}| \psi\>
=\<\psi|{\hat M}_{I}\hat{B}_I^{\dagger}|\psi\>
=0 \ , \nonumber\\
& &\< \psi|{\hat M}_{O}\hat{B}_O| \psi\>
=\<\psi|\hat{B}_O^{\dagger}{\hat M}_{O}| \psi\>
=\< \psi|\hat{B}_O {\hat M}_{O}| \psi\>
=\<\psi|{\hat M}_{O}\hat{B}_O^{\dagger}|\psi\>
=0 \ , \label{a-1}\\
& &\< \psi|\hat{B}_I| \psi\>
=\<\psi|\hat{B}_I^{\dagger}| \psi\>
=\< \psi|\hat{B}_O| \psi\>
=\<\psi|\hat{B}_O^{\dagger}| \psi\>
=0 \ , 
\label{a-2}\\
& &
\< \psi|\hat{M}_{I}| \psi\>
=4|\beta_I|^{2}\frac{\Psi'_{I}}{\Psi_{I}},~~~~
\< \psi|\hat{M}_{O}| \psi\>
=4|\beta_O|^{2}\frac{\Psi_{O}'}{\Psi_{O}} \ , 
\label{a-3}\\
& &
\< \psi|\hat{M}_{I}^2| \psi\>
=16|\beta_I|^{2}\frac{\Psi'_{I}}{\Psi_{I}}
+16|\beta_I|^{4}\frac{\Psi''_{I}}{\Psi_{I}}\ 
,~~~~\nonumber\\
& &\< \psi|\hat{M}_{O}^2| \psi\>
=16|\beta_O|^{2}\frac{\Psi_{O}}{\Psi_{O}}
+16|\beta_O|^{4}\frac{{\Psi_O}\!''}{\Psi_{O}} \ . 
\label{a-4}\\
& &\< \psi|\hat{B}_I^{\dagger}\hat{B}_I| \psi\>
=2|\beta_I|^{2}
\left(
1-\frac{1}{2S_{I}}
\right)
\frac{\Psi'_{I}}{\Psi_{I}}
-\frac{4|\beta_I|^{4}}{2S_I}
\frac{\Psi''_{I}}{\Psi_{I}}
,~~~~\nonumber\\
& &\< \psi|\hat{B}_O^{\dagger}\hat{B}_O| \psi\>=2|\beta_O|^{2}
\left(
1-\frac{1}{2S_{O}}
\right)\frac{\Psi'_{O}}{\Psi_{O}}
-
\frac{4|\beta_O|^{4}}{2S_{O}}
\frac{\Psi''_{O}}{\Psi_{O}} \ , 
\label{a-5}\\
& &
\< \psi|{\hat B}_I{\hat B}_I^{\dagger}| \psi\>=
\< \psi|{\hat B}_I^{\dagger}{\hat B}_I| \psi\>+
\left(
1-\frac{\< \psi|{\hat M}_{I}| \psi\>}{2S_{I}}
\right)
,~~~~\nonumber\\
& &
\< \psi|{\hat B}_O{\hat B}_O^{\dagger}| \psi\>=
\< \psi|{\hat B}_O^{\dagger}{\hat B}_O| \psi\>+
\left(
1-\frac{\< \psi|M_{O}| \psi\>}{2S_{O}}
\right) \ , 
\label{a-6}\\
& &
\bra{\psi}{\hat B}_I^2\ket{\psi}=2\beta_I\frac{\Psi_I'}{\Psi_I} \ , \qquad
\bra{\psi}{\hat B}_I^{\dagger 2}\ket{\psi}
=2\beta_I^*\frac{\Psi_I'}{\Psi_I} \ , \nonumber\\
& &
\bra{\psi}{\hat B}_O^2\ket{\psi}=2\beta_O\frac{\Psi_O'}{\Psi_O} \ , \qquad
\bra{\psi}{\hat B}_O^{\dagger 2}\ket{\psi}
=2\beta_O^*\frac{\Psi_O'}{\Psi_O} \ . 
\label{a-7}
\end{eqnarray}
Thus, the expectation value of the Hamiltonian in (\ref{5-1}) can be 
calculated by using the above expectation values.

\section{Various expectation values by using the fermionic squeezed state}

In this appendix, we summarize the various expectation values with respect 
to the fermionic squeezed state developed in \S 5-3. 
\begin{eqnarray}
& &\< s|{\hat M}_{I}| s\>
=\frac{1}{\Gamma}
\Biggl\{~\!
\sum_{n=0}^{[\frac{\Omega}{2}]}
\frac{[(2n)!]^2}{(n!)^{2}}
~\!\sum_{r=0}^{2n}
|\beta_I|^{2r}~\!|\beta_O|^{4n-2r}
\frac{2r}{r!~\!(2n-r)!}
\nonumber\\
& &\qquad\qquad\qquad\qquad
\times
\prod^{r-1}_{p=1}
\left(
1-\frac{p}{2S_{I}}
\right)
\prod^{(2n-r-1)}_{q=1}
\left(
1-\frac{q}{2S_{O}}
\right)
\Biggr\}\ ,  \nonumber\\
& &\< s|{\hat M}_{O}| s\>
=
\frac{1}{\Gamma}
\Biggl\{~\!
\sum_{n=0}^{[\frac{\Omega}{2}]}
\frac{[(2n)!]^2}{(n!)^{2}}
~\!\sum_{r=0}^{2n}
|\beta_O|^{2r}~\!|\beta_I|^{4n-2r}
\frac{2r}{r!~\!(2n-r)!}\nonumber\\
& &\qquad\qquad\qquad\qquad
\times 
\prod^{r-1}_{p=1}
\left(
1-\frac{p}{2S_{O}}
\right)
\prod^{(2n-r-1)}_{q=1}
\left(
1-\frac{q}{2S_{I}}
\right)
\Biggr\} \ , \nonumber\\
& &\label{b-1}\\
& &\< s|{\hat M}_{I}^{2}| s\>
=
\frac{1}{\Gamma}
\Biggl\{~\!
\sum_{n=0}^{[\frac{\Omega}{2}]}
\frac{[(2n)!]^2}
{(n!)^{2}}
~\!\sum_{r=0}^{2n}
|\beta_I|^{2r}~\!|\beta_O|^{4n-2r}
\left(
1-\frac{p}{2S_{I}}
\right)
\nonumber\\
& &\qquad\qquad\qquad
\times
\prod^{(2n-r-1)}_{q=1}
\left(
1-\frac{q}{2S_{O}}
\right)
\Biggr\} \ , \nonumber\\
& &\< s|{\hat M}_{O}^{2}| s\>
=\frac{1}{\Gamma}
\Biggl\{~\!
\sum_{n=0}^{[\frac{\Omega}{2}]}
\frac{[(2n)!]^2}
{(n!)^{2}}
~\!\sum_{r=0}^{2n}
|\beta_O|^{2r}~\!|\beta_I|^{4n-2r}\frac{4r^{2}}{r!~\!(2n-r)!}
\nonumber\\
& &\qquad\qquad\qquad\qquad
\times
\prod^{r-1}_{p=1}
\left(
1-\frac{p}{2S_{O}}
\right)
\prod^{(2n-r-1)}_{q=1}
\left(
1-\frac{q}{2S_{I}}
\right)
\Biggr\} \ , 
\nonumber\\
& &\< s|{\hat M}_{I}{\hat M}_{O}| s\>
=\< s|{\hat M}_{O}{\hat M}_{I}| s\>\nonumber\\
& &\ \ =\frac{1}{\Gamma}
\Biggl\{~\!
\sum_{n=0}^{[\frac{\Omega}{2}]}
\frac{[(2n)!]^2}{(n!)^2}
\sum_{r=0}^{2n}
|\beta_I|^{2r}~\!|\beta_O|^{4n-2r}
\frac{4r~\!(2n-r)}{r!~\!(2n-r)!}~\!
\nonumber\\
& &\qquad\qquad\qquad\qquad
\times
\prod^{r-1}_{p=1}
\left(
1-\frac{p}{2S_{I}}
\right)
\prod^{2n-r-1}_{q=1}
\left(
1-\frac{q}{2S_{O}}
\right)
\Biggr\} \ , \nonumber\\
& &\label{b-2}\\
& &\< s|{\hat B}_I{\hat B}_I^{\dagger}| s\>=
\frac{1}{\Gamma}
\Biggl\{~\!\sum^{[\frac{\Omega}{2}]}_{n=0}
\frac{[(2n)!]^{2}}{(n!)^2}
\sum_{r=0}^{2n}
|\beta_I|^{2r}~\!|\beta_O|^{4n-2r}
\frac{(r+1)}{r!~\!(2n-r)!}\nonumber\\
& &\qquad\qquad\qquad\qquad
\times
\prod^{r}_{p=1}
\left(
1-\frac{p}{2S_{I}}
\right)
\prod^{2n-r-1}_{q=1}
\left(
1-\frac{q}{2S_{O}}
\right)\Biggr\} \ , \nonumber\\
& &\< s|{\hat B}_O{\hat B}_O^{\dagger}| s\>=
\frac{1}{\Gamma}
\Biggl\{~\!\sum^{[\frac{\Omega}{2}]}_{n=0}
\frac{[(2n)!]^{2}}{(n!)^2}
\sum_{r=0}^{2n}
|\beta_O|^{2r}~\!|\beta_I|^{4n-2r}
\frac{(r+1)}{r!~\!(2n-r)!}
\nonumber\\
& &\qquad\qquad\qquad\qquad
\times
\prod^{r}_{p=1}
\left(
1-\frac{p}{2S_{O}}
\right)
\prod^{2n-r-1}_{q=1}
\left(
1-\frac{q}{2S_{I}}
\right)\Biggr\} \ , \nonumber\\
& &\label{b-3}\\
& &\< s|{\hat B}_I^{\dagger 2}| s\>=\frac{1}{\Gamma}
\Biggl\{~\!
\sum_{n=0}^{[\frac{\Omega}{2}]}
\frac{(2n)!~\!(2n+2)!}{n! \!~(n+1)!}
\sum_{r=0}^{2n}
\frac{\beta_I^2|\beta_I|^{2r}~\!|\beta_O|^{4n-2r}}{r!~\!(2n-r)!}
\nonumber\\
& &\qquad\qquad\qquad\qquad
\times
\prod^{r+1}_{p=1}
\left(
1-\frac{p}{2S_{I}}
\right)
\prod^{(2n-r-1)}_{q=1}
\left(
1-\frac{q}{2S_{O}}
\right)
\Biggr\}\ , \nonumber\\
& &\< s|B_O^{\dagger 2}| s\>=\frac{1}{\Gamma}
\Biggl\{~\!
\sum_{n=0}^{[\frac{\Omega}{2}]}
\frac{(2n)!~\!(2n+2)!}{n! \!~(n+1)!}
\sum_{r=0}^{2n}
\frac{\beta_O^2|\beta_O|^{2r}~\!|\beta_I|^{4n-2r}}{r!~\!(2n-r)!}
\nonumber\\
& &\qquad\qquad\qquad\qquad
\times
\prod^{r+1}_{p=1}
\left(
1-\frac{p}{2S_{O}}
\right)
\prod^{(2n-r-1)}_{q=1}
\left(
1-\frac{q}{2S_{I}}
\right)
\Biggr\} \ , \nonumber\\
& &\label{b-4}\\
& &\< s|{\hat B}_I{\hat B}_O| s\>
=\frac{1}{\Gamma}
\Biggl\{~\!
\sum^{[\frac{\Omega}{4}]}_{n=0}
\frac{(2n)!(2n+2)!}{n!~\!(n+1)!}
\sum^{2n}_{r=0}
\frac{\beta_I \beta_O|\beta_I|^{2r}~\!|\beta_O|^{4n-2r}}{r!~\!(2n-r)!}
\nonumber\\
& &\qquad\qquad\qquad\qquad
\times
\prod^{r}_{p=1}
\left(
1-\frac{p}{2S_{I}}
\right)
\prod^{2n-r}_{q=1}
\left(
1-\frac{q}{2S_{O}}
\right)\Biggr\} \ , \nonumber\\
& &\< s|{\hat B}_I^{\dagger}{\hat B}_O| s\>=
\frac{1}{\Gamma}
\Biggl\{~\!
\sum^{[\frac{\Omega}{2}]}_{n=0}
\frac{[(2n)!]^2}{(n!)^2}
\sum_{r=0}^{2n}
\beta_I^* \beta_O |\beta_I|^{2r}~\!|\beta_O|^{4n-2r-2}
~\!
\frac{(2n-r)}{r!~\!(2n-r)!} \nonumber\\
& &\qquad\qquad\qquad\qquad
\times
\prod^{r}_{p=1}
\left(
1-\frac{p}{2S_{I}}
\right)
\prod^{2n-r-1}_{q=1}
\left(
1-\frac{q}{2S_{O}}
\right)~\!\Biggl\}\ . 
\label{b-5}
\end{eqnarray}
The expectation values of other products of two operators,
except for the complex conjugates of the above operators, 
are zero.

\end{document}